# Micro-explosion of emulsion droplets with nanoparticles at high temperature

Houpeng Zhang[1], Zhen Lu[1], Tianyou Wang[1,2], Zhizhao Che[1,2,*]

1. State Key Laboratory of Engine, Tianjin University, Tianjin, 300350, China

2. National Industry-Education Platform of Energy Storage, Tianjin University, Tianjin, 300350, China

*Corresponding author: chezhizhao@tju.edu.cn

## Abstract

Compared with traditional fuels, emulsified fuels can improve fuel atomization and combustion, and nanoparticles as additives have the potential to enhance combustion and reduce emissions. Previous studies on micro-explosion mainly considered emulsion droplets, but the role of nanoparticles in emulsion droplets is still unclear. In this study, we experimentally investigate the micro-explosion of emulsion droplets with nanoparticles via high-speed photography, digital image processing, optical microscopy, and scanning electron microscopy. The results show that the presence of nanoparticles can greatly improve the strength and probability of micro-explosion, particularly for carbon nanoparticles. This is mainly because nanoparticles can agglomerate during the evaporation of emulsion droplets, facilitate the absorption of radiation energy, inhibit the diffusion of superheated vapor, and ultimately promote micro-explosion. The effects of nanoparticle mass fraction and water content are also investigated, and the results show that the increase of nanoparticles and water can facilitate micro-explosion.

**Keywords**: Droplet micro-explosion; emulsion droplets; Multi-component fuels; Nanoparticles.

## 1. Introduction

The efficient atomization and combustion of liquid fuels is a major problem in the process of energy conversation and utilization [1,2]. Due to the emission problems faced by traditional fuels such as diesel and gasoline, seeking alternative clean fuels has become the focus of attention. In recent years, various liquid fuels have been tested in many engine experiments [3], and emulsified fuel is considered to be an alternative clean fuel with high combustion efficiency and low pollutant emissions [4-6].

### 1.1 Emulsified fuels

Emulsified fuels have been tested in numerous combustion systems to control flames, improve combustion efficiency, and reduce NOx emissions [7-9]. Emulsified fuels can lead to puffing and micro-explosion in the spray combustion process [10-12]. During micro-explosion, the lighter components of the emulsion droplets vaporize instantaneously, and the entire droplet breaks into many small droplets, thereby improving the atomization efficiency of the fuels [13]. However, during the combustion of water-oil emulsified fuels, due to the quick vaporization of water, which has a lower boiling point than oil but a high latent heat, the combustion temperature is also reduced compared to the neat fuel and the fuel consumption is increased [14]. Therefore, to utilize the advantages of emulsified fuels to reduce NOx emissions and, meanwhile, overcome the increase in fuel consumption caused by water, adding nanoparticles to emulsified fuels has become an effective strategy [15].

## 1.2 Nanofluid fuels

Nanofluid fuels can be prepared by adding nanoparticles (mostly between 10-100 nm) to liquid fuels at a certain proportion [16]. Nanofluid fuels have high energy density and high thermal conductivity, which can shorten the ignition delay time during combustion, thereby increasing the combustion and heat release rate, and reducing emissions of various pollutants [6,16-19]. Studies have shown that diesel nanofluids with metal oxide additives can significantly increase combustion performance and decrease pollutant emissions [20]. Nanofluid fuels with $Fe_3O_4$ nanoparticles can improve combustion characteristics, even at very low concentrations, while reducing $NO_x$ and $SO_2$ emissions [21]. A small amount of cerium-zirconium oxide nanoparticles can also improve the braking thermal efficiency of diesel engines and reduce soot emissions [22]. Cerium oxide nanoparticles can improve the effective thermal efficiency of diesel combustion and reduce the emissions of HC, CO, $NO_x$, and soot [23]. Alumina nanoparticles and carbon nanotube nanoparticles can reduce the emission of $NO_x$ and smoke compared with neat biodiesel [24].

In addition to affecting thermal efficiency and pollutant emission, nanoparticles in nanofluid fuels can improve atomization by promoting evaporation and micro-explosion. Nanoparticles can absorb thermal radiation, improve the heat transfer inside the droplets, and accelerate the evaporation of nanofluid fuels [25-27]. In addition, the agglomeration of nanoparticles on the droplet surface inhibits the vapor transport in the nanofluid fuel droplets [28]. Hence, an important phenomenon is that nanoparticles can cause micro-explosion of nanofluid fuel droplets, resulting in severe breakup of nanofluid fuel droplets and promoting atomization [29,30].

Many experiments have been performed to study the evaporation and micro-explosion of nanofluid fuel droplets [31,32]. Wang *et al.* [28] analyzed the evaporation/micro-explosion process of nanofluid fuel droplets with cerium oxide nanoparticles and found that increasing the nanoparticle concentration can reduce the micro-explosion delay. Wang *et al*. [33] found that the addition of cerium oxide can cause repeated intense micro-explosion of the fuel droplets. Ashok *et al*. [34] studied the effects of zinc oxide particles in biodiesel on the performance of direct injection engines, and found that the catalytic effect and the micro-explosion-promotion effect of nanoparticles are the main reasons for the improvement of combustion performance and braking thermal efficiency.

Many studies have focused on the influencing factors of nanofluid fuel droplets on micro-explosion, such as the heating temperature, and the concentration, type, and size of nanoparticles. Tanvir *et al*. [35] found that the evaporation of ethanol droplets with graphite nanoparticles is faster than pure ethanol droplets in the early stage, but lower in the later stage. This may be because the absorption of radiation energy by particles is dominant in the early stage, and particle agglomeration is dominant in the later stage and reduces the effective evaporation surface area of droplets. Ferrao *et al*. [36] experimentally found that increasing the aluminum nanoparticle concentration can increase the micro-explosion intensity and decrease the micro-explosion delay time. This is because the nanoparticles enhance the heterogeneous nucleation, produce more bubbles, and micro-explosion occurs when bubbles aggregate. Ghamari *et al*. [37] investigated the influence of carbon nanoparticles, nanotubes, and nanoplatelets on the evaporation of aviation fuels, and found that carbon nanotubes have an intense effect on promoting evaporation.

### 1.3 Emulsified fuels with nanoparticles

Water-oil emulsified fuel with nanoparticles is a new type of clean alternative fuel, which can improve effective fuel utilization by promoting micro-explosion and reduce $NO_x$ emissions. Dhinesh *et al*. [38] used cerium oxide nanoparticles in an emulsion of biofuel and found that, compared with ordinary emulsified fuels, nanofluid emulsion droplets cause micro-explosion, which increases fuel evaporation and reduces smoke emissions. Basha *et al*. [39] used carbon nanotubes (CNT) mixed with water-diesel emulsion droplets to study the performance and emissions of diesel engines. Because the water droplets wrapped by CNT can quickly absorb heat at high temperature, water vapor break through the surrounding oil film and cause micro-explosion, resulting in many small droplets fully reacting with diesel and air and enhancing the catalytic combustion of CNT, thus improving the engine efficiency. E *et al.* [40] studied the effects of biodiesel-diesel blends with water and cerium oxide nanoparticles on the combustion emissions and the performance of marine engines, and found that the presence of cerium oxide $CeO_2$ nanoparticles and the occurrence of micro-explosion significantly improve the combustion characteristics, thereby reducing the effective fuel consumption and CO, PM, $NO_x$, and HC emissions. Despite the aforementioned studies in this field, an in-depth understanding of the micro-explosion of emulsified fuels with nanoparticles is still needed for their applications.

In summary, emulsified fuel with nanoparticles has been considered a clean alternative fuel to reduce the pollutant emissions of combustion systems. An important advantage of emulsion fuel with nanoparticles is that it can induce puffing and micro-explosion and promote secondary breakup, thereby improving the atomization of liquid fuels and reducing emissions of pollutants. At present, the existing studies on emulsified fuels with nanoparticles mainly focus on their combustion characteristics, emission products, and combustion efficiency under internal combustion engine conditions. Understanding the micro-explosion mechanism of single emulsion droplets with nanoparticles is important for spray combustion applications. In the present study, the dynamic process of micro-explosion of emulsion droplets with nanoparticles is investigated experimentally. The details of micro-explosion are imaged by high-speed photography, and analyzed by digital image processing. The microscopic morphology of emulsion droplets with nanoparticles is observed under optical and scanning electron microscopes. The absorbance of emulsion droplets with different nanoparticles is measured by a spectrophotometer. The intensity and probability of micro-explosion are quantified, and the influences of nanoparticle mass fractions and water content are also studied.

## 2. Experimental

### 2.1 Preparation and characterization of emulsion droplets with nanoparticles

#### 2.1.1 Microstructure of nanoparticles

In this study, alumina and carbon nanoparticles with an average particle size of about 50 nm are selected. Their microstructure was measured by scanning electron microscopy (SEM, Regulus 8100), as shown in Figure 1. The images show that both alumina and carbon nanoparticles exist in the form of small spheres of about 50 nm. There is obvious agglomeration between the particles, and the size of the agglomeration is about 2--10 μm.

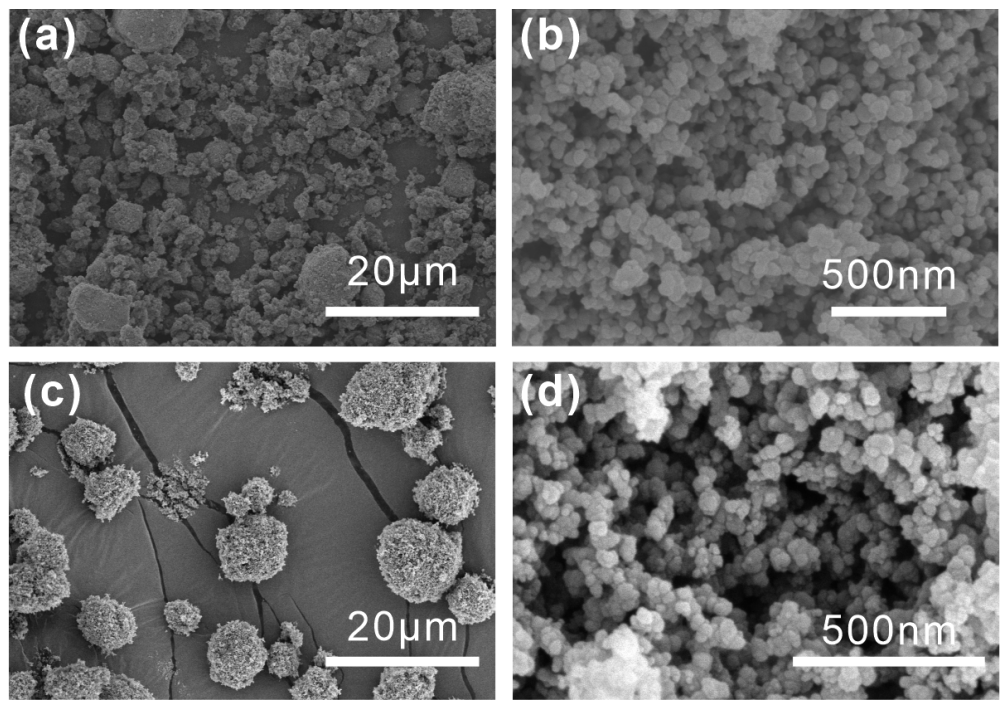

Figure 1. SEM images of nanoparticles: (a, b) alumina nanoparticles, (c, d) carbon nanoparticles.

Table 1. Physical properties of nanoparticles.

| Properties | Alumina nanoparticles | Carbon nanoparticles |
|---|---|---|
| Density, $\rho_s$ [$\mathrm{kg/m^3}$] | 3980 | 2300 |
| Specific heat capacity, $c_p$ [$\mathrm{kJ/(kg \cdot K)}$] | 0.78 | 0.502 |
| Thermal conductivity, $k$ [$\mathrm{W/(m \cdot K)}$] | 30 | 2300 |
| Particle diameter, $d_s$ [nm] | 50 | 50 |

### 2.1.2 Preparation of emulsified fuel with nanoparticles

In the preparation of emulsified fuels with nanoparticles, we used commercial diesel as the base fuel and Span 80 as the surfactant. Alumina and carbon nanoparticles of 50 nm are selected, and their physical parameters are shown in Table 1. We first used an electronic balance (QUINT IX224-1CN) to weigh the mass of alumina and carbon nanoparticles. The nanoparticles were added to deionized water and stirred at 3000 rpm for about 10 minutes using a magnetic stirrer. Then the surfactant Span 80 and the mixture of nanoparticles and water were added to the diesel, and mixed for 30 minutes by a magnetic stirrer to stabilize the nanoparticles in the emulsion. The emulsified fuels with nanoparticles are stable for at least half an hour after the preparation, hence the experiments were carried out within this time. The mass fraction of the nanoparticles is defined as $w_{\mathrm{NP}} = m_{\mathrm{NP}} / m_{\mathrm{total}}$, where $m_{\mathrm{NP}}$ is the mass of the nanoparticles and $m_{\mathrm{total}}$ is the total mass of the emulsified fuel with nanoparticles. $w_{\mathrm{NP}}$ was varied in the range of 0.05--3.0 wt% in the experiment. The low particle concentration can avoid stratification of the nanofluid fuel, which can occur if the particle concentration is too high. The water content in the emulsified fuel is defined as $\phi_{\mathrm{water}} = V_{\mathrm{water}} / \left( V_{\mathrm{water}} + V_{\mathrm{oil}} + V_{\mathrm{surfactant}} \right)$, where $V_{\mathrm{water}}$, $V_{\mathrm{oil}}$, and $V_{\mathrm{surfactant}}$ are the volumes of water, oil, and surfactant in preparing the emulsified fuel, respectively. $\phi_{\mathrm{water}}$ was varied in the range of 10--40 vol%. The surfactant content in the emulsified fuel is defined as $\phi_{\mathrm{surfactant}} = V_{\mathrm{surfactant}} / \left( V_{\mathrm{water}} + V_{\mathrm{oil}} + V_{\mathrm{surfactant}} \right)$, and was fixed at 1 vol%.

### 2.1.3 Spectral absorption characteristics of the emulsified fuels with nanoparticles

To study the absorption of infrared radiation by emulsified fuels with nanoparticles, we used a spectrophotometer (LAMBDA 750) to measure the transmission spectrum of the emulsified fuels with nanoparticles in the near-infrared wavelength range (800--1100 nm). According to the Lambert-Beer law:

$$A = \lg \frac{I_0}{I} = -\lg T \tag{1}$$

where $A$ is the absorbance, $T$ is the transmittance, and $I_0$ and $I$ are the intensities of the incident light and transmitted light, respectively.

According to the measured transmission spectrum in the near-infrared wavelength range, the absorbance $A$ can be obtained. The absorbance curve of the emulsified fuels at near-infrared wavelengths (800-1100 nm) is shown in Figure 2. In the near-infrared band, the absorbance of the emulsified fuel with 0.5 wt% carbon nanoparticles is much higher than that of the emulsified fuel with 0.5 wt% alumina nanoparticles, which indicates that, in the heating process, the emulsified fuel with carbon nanoparticles absorbs more radiation energy to increase the droplet temperature. The absorbance of the emulsified fuel with alumina nanoparticles is slightly higher than that without nanoparticles, indicating that the emulsified fuels with alumina nanoparticles have a slightly higher ability to absorb radiation heat than the neat emulsified fuel.

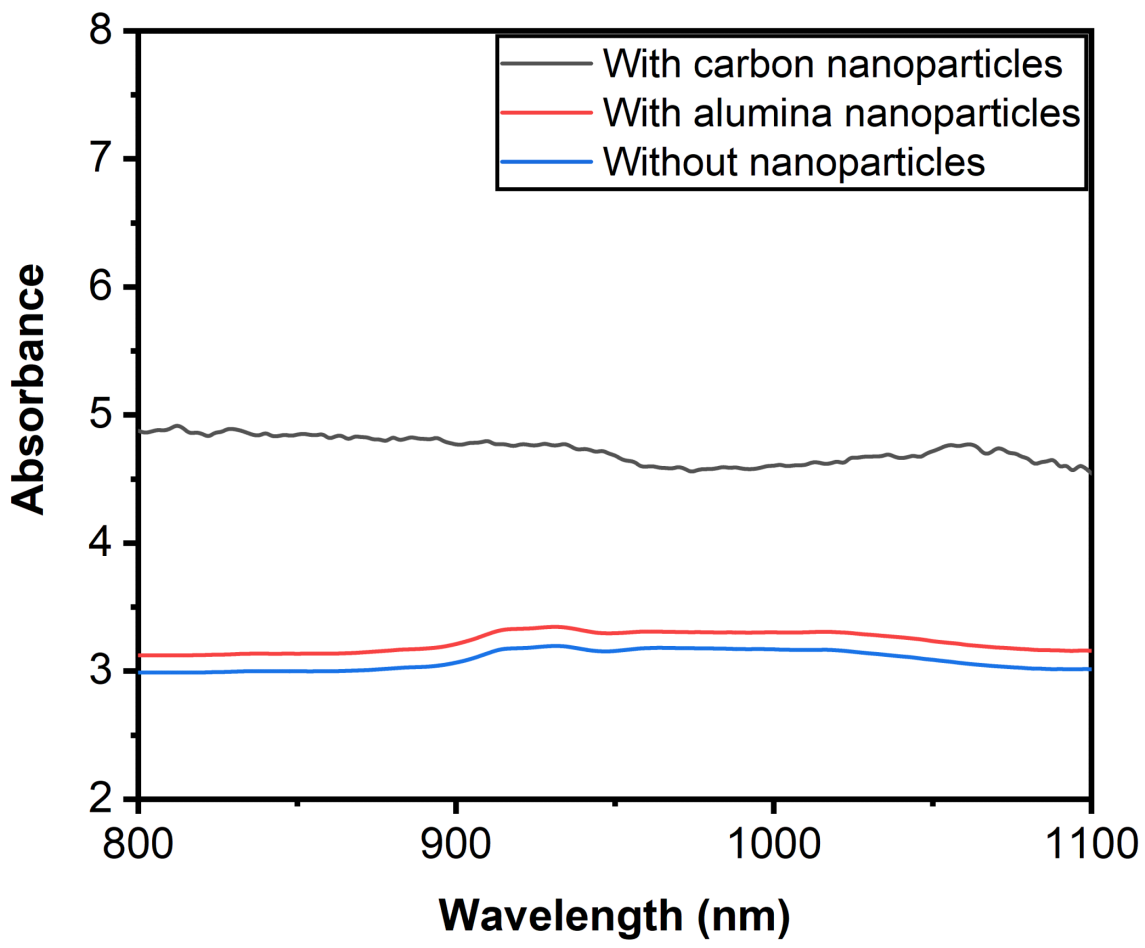

Figure 2. Absorbance of the emulsified fuels against the wavelength. Here, $w_{NP}$ = 0.5 wt%.

## 2.2 Experimental setup

The experimental setup is similar to our previous study for neat emulsion droplets [41], as shown in Figure 3, and it is only described briefly here. Emulsion droplets were suspended at a thermocouple junction (wire diameter 120 μm), which was also used to monitor the transient temperature of the droplet. The droplet was heated by a continuous-wave infrared laser (1064 nm) at its maximum power (about 2 W). The droplet evaporation and breakup process was captured by a high-speed camera (Photron SA1.1) at 8000 fps with a short exposure time of 2 μs. A macrolens with a small aperture setting (F22) was used for the imaging. A high-power LED light was used for illumination. A notch filter (1064 nm) was added in front of the macrolens to eliminate the infrared laser light entering the camera. To ensure consistency between different experiments, the same experimental settings were used in all experiments except the variable that we studied its effects. To further know the uncertainty of the mist concentration and ensure the reliability of the results, we repeated the experiment about 30 times for each condition.

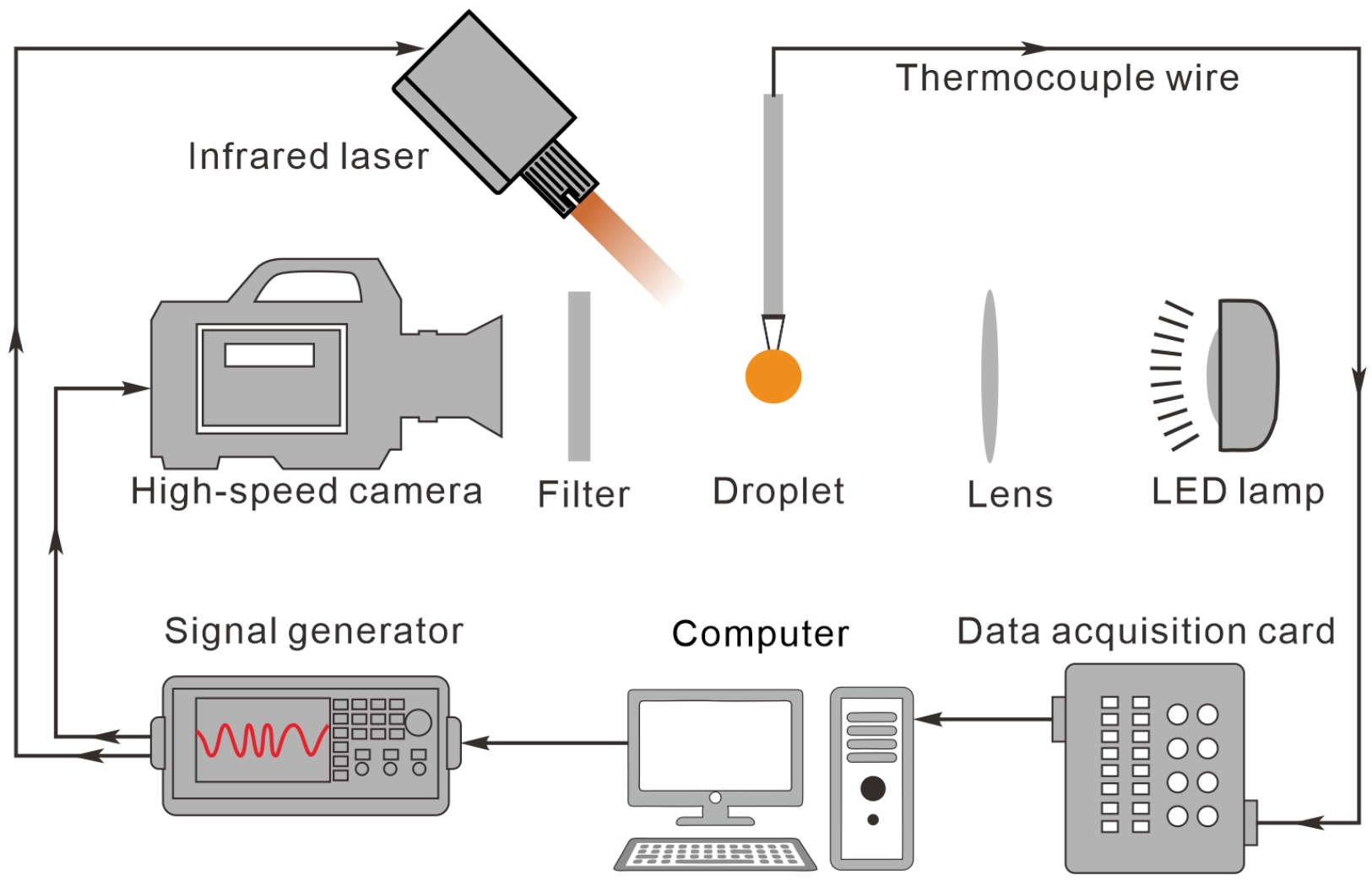

Figure 3. Experimental setup for the visualization of droplet micro-explosion.

### 2.3 Morphology of nanoparticles in emulsion droplets

To obtain the microscopic state of nanoparticles in emulsion droplets, the evaporation process of emulsion droplets with nanoparticles was observed under an optical microscope. Firstly, the emulsion droplets with nanoparticles were deposited on a thin glass slide, and an LED light source (Hecho S5000) was used to illuminate directly above the glass slide. A CMOS camera (FLIR GS3-U3-123S6M-C) and an inverted microscope (Nikon Ti-U) were used to photograph the evaporation process of the emulsion droplets with nanoparticles from the bottom. Due to the thin liquid film on the glass slide, the dispersed water phase in the emulsion droplets evaporates quickly under strong light irradiation.

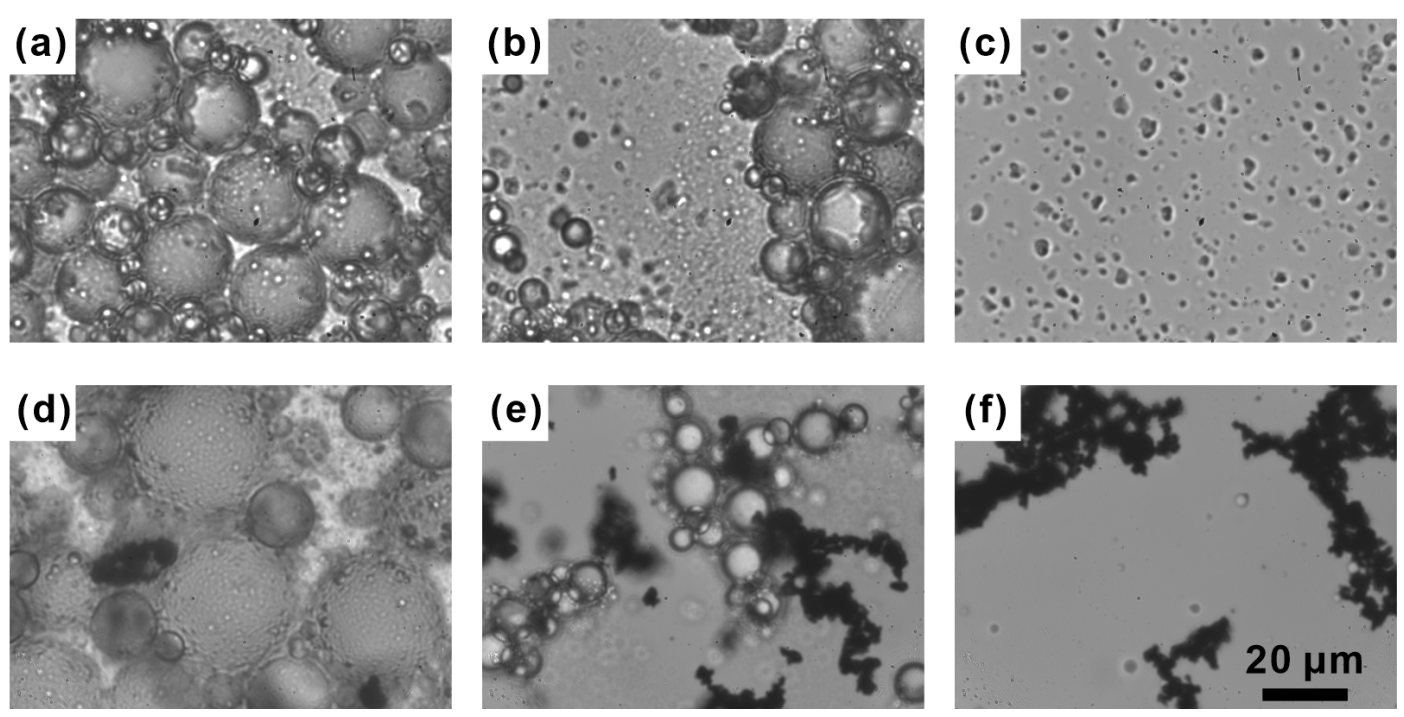

Figure 4. Microscopic process of the evaporation of emulsion droplets with nanoparticles. (a, b, c) alumina nanoparticles, (d, e, f) carbon nanoparticles. The times of the images taken are: (a, d) just deposited on the glass, (b, e) during the evaporation, and (c, f) after the complete evaporation.

The evaporation processes of the emulsion droplets with nanoparticles are shown in Figure 4. For the emulsion droplet with alumina nanoparticles, when the emulsion droplet is just deposited, many nanoparticles can be observed suspended at the oil-water interface, as shown in Figure 4a. As the evaporation proceeds, the dispersed water droplets in the emulsion droplet gradually evaporate, and more nanoparticles gradually appear, as shown in Figure 4b. When the dispersed water droplets are completely evaporated, particle agglomerations can be observed clearly, as shown in Figure 4c. For the emulsion droplets with carbon nanoparticles, there are obvious particles at the oil-water interface at the initial moment, as shown in Figure 4d. As the evaporation process proceeds, the carbon nanoparticles form many dendritic agglomerations, as shown in Figure 4e. Especially when the evaporation completes, the nanoparticles mainly exist in the form of large agglomerations, and the agglomeration size is much larger than that of alumina, as shown in Figure 4f.

To quantitatively characterize the agglomeration size in the late stage of the evaporation of emulsion droplets, the size and number of agglomerates are counted by using a customized image processing program (see Section S1 in the Supplementary Material for details). For emulsion droplets with alumina nanoparticles, 20 groups of images at different positions in the late stage of evaporation were selected for agglomerate size statistics. For emulsion droplets with carbon nanoparticles, due to the large agglomerates, the number of agglomerates in each image is small. Hence, 50 groups of images at different positions in the late stage of evaporation were selected for agglomerate size statistics. The statistical results are shown in Figure 5. For emulsion droplets with alumina nanoparticles, the size of the alumina nanoparticles is mainly distributed in the range of 0--3 μm, which accounts for about 71% of all agglomerates. For emulsion droplets with carbon nanoparticles, the size of carbon nanoparticle agglomerates is mainly distributed in the range of 0--50 μm, which accounts for about 68% of all agglomerates. The above results show that the agglomerates formed by carbon nanoparticles after evaporation are larger and easier to agglomerate than alumina nanoparticles.

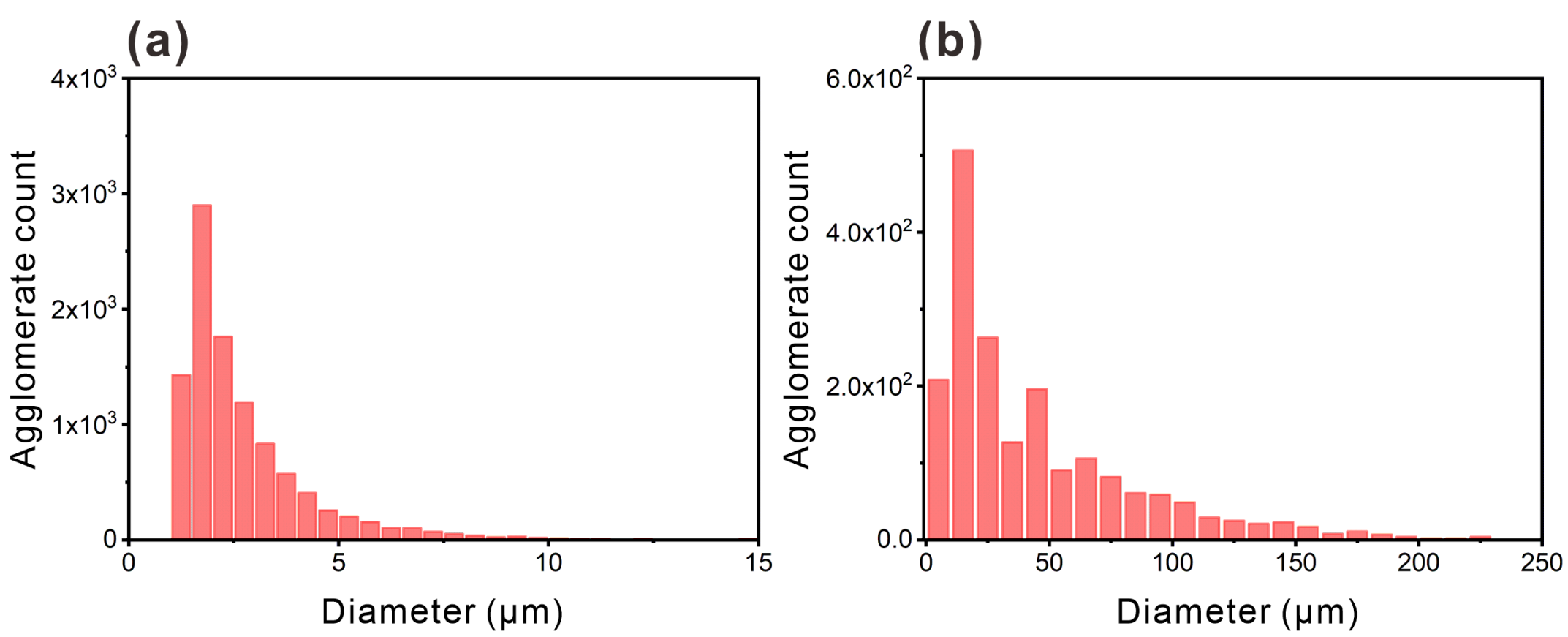

Figure 5. Size distribution of particle agglomerates after the evaporation of the emulsion droplets with nanoparticles: (a) alumina nanoparticles, (b) carbon nanoparticles. Here, $w_{NP}$ = 0.5 wt%.

# 3. Results and discussion

## 3.1 Effect of nanoparticles on micro-explosion of emulsion droplets

Firstly, the influence of the presence of nanoparticles on the micro-explosion process is compared, as shown in Figure 6. The micro-explosion processes with alumina and carbon nanoparticles are shown in Figures 6b and 6c. For comparison, the micro-explosion processes without nanoparticles are shown in Figure 6a. For the emulsion droplet without nanoparticles, as the droplets are continuously heated, the emulsion droplets can break up at a certain moment. During the breakup process, the liquid film is completely stretched until it breaks up into many secondary droplets, which are ejected around at a high speed. At the moment of micro-explosion, some liquid vaporizes suddenly, and the violent airflow results in a rapid dispersion of the vapor, so the vapor temperature decreases rapidly, forming mist [41]. For the emulsion droplet with alumina nanoparticles, the micro-explosion process is more intense than that of the emulsion droplet without nanoparticles, and the number of secondary droplets produced by its fragmentation is larger and the size is smaller. For the emulsion droplet with carbon nanoparticles, the breakup process is very intense during the heating process. The liquid of the emulsion droplet is completely consumed in a very short period (less than 3 ms, as shown in Figure 6c). Hence, the expansion and stretching of the liquid film are not observed, and the entire droplet instantaneously breaks up into many small droplets. In addition, the emulsion droplets with carbon nanoparticles also produce much mist throughout the micro-explosion process, and the mist concentration is significantly higher than that for alumina nanoparticles. Therefore, the presence of nanoparticles makes the micro-explosion more intense, particularly for carbon nanoparticles.

The influence of nanoparticles is also reflected in the temperature change of droplets during heating and micro-explosion. The micro-explosion process of droplets brings drastic temperature changes. The transient temperature of the emulsion droplet during heating and breakup was measured by the thermocouple. The temperature changes of the three emulsion droplets during evaporation and micro-explosion are presented in Figure 7. It can be seen that the temperature change of emulsion droplets has three stages. First, the temperature of the droplets heated by the infrared laser rises rapidly. When the temperature of the droplets exceeds 100 °C, the evaporation of the droplet intensifies, resulting in a slower increase in the droplet temperature. When the droplet explodes, the droplet temperature decreases suddenly, although the heating effect of the laser is still ongoing. The significant decrease in temperature is due to the sudden vaporization of the water during the micro-explosion and absorption of latent heat of vaporization [41].

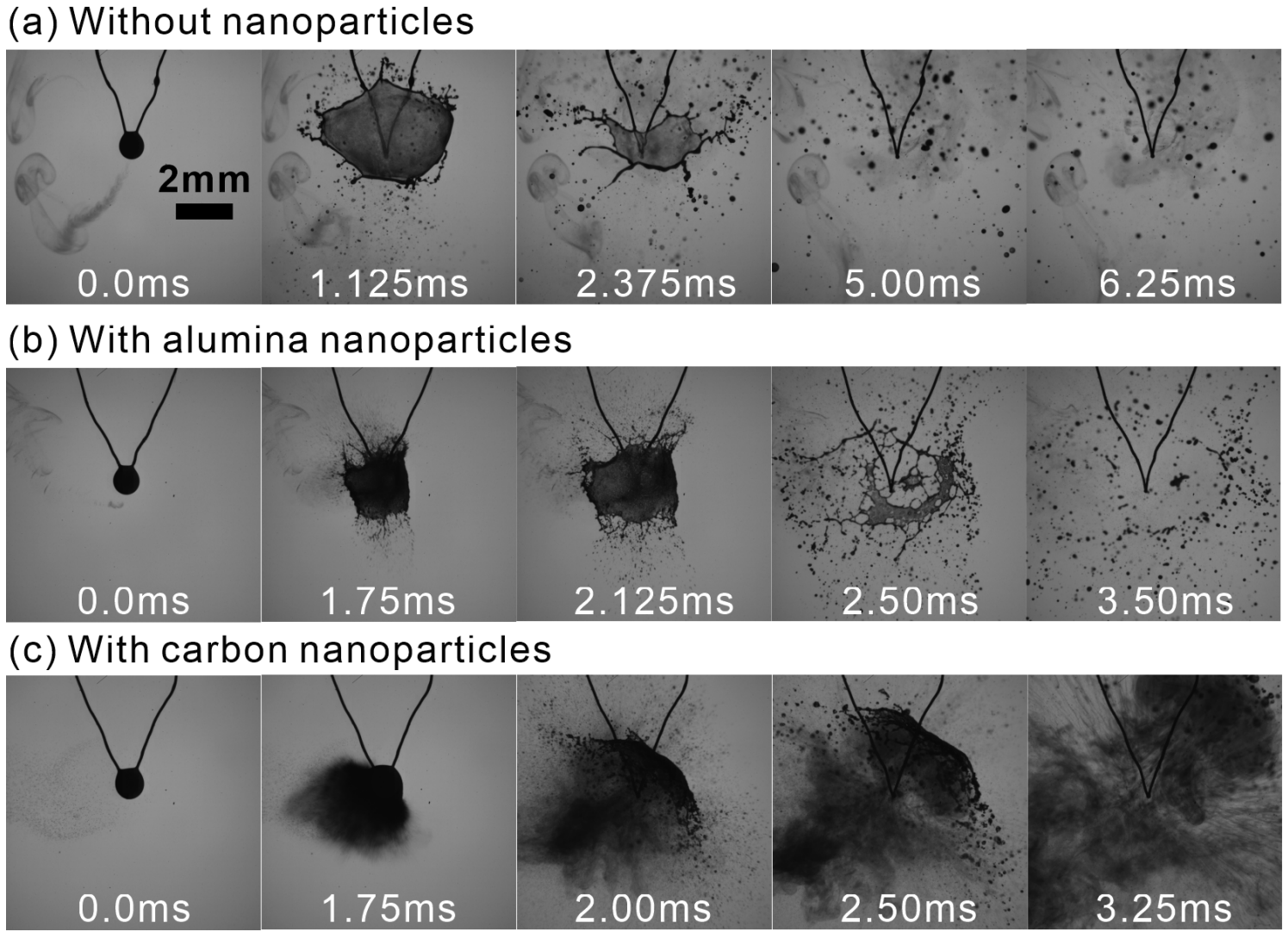

Figure 6. Micro-explosion processes of emulsion droplets with/without nanoparticles: (a) without nanoparticles, (b) with alumina nanoparticles, (c) with carbon nanoparticles. Here, $w_{NP}$ = 0.5 wt%, and $\phi_{water}$ = 30 vol%. Video clips for the three micro-explosion processes are provided as Movies 1--3 in Supplementary Materials.

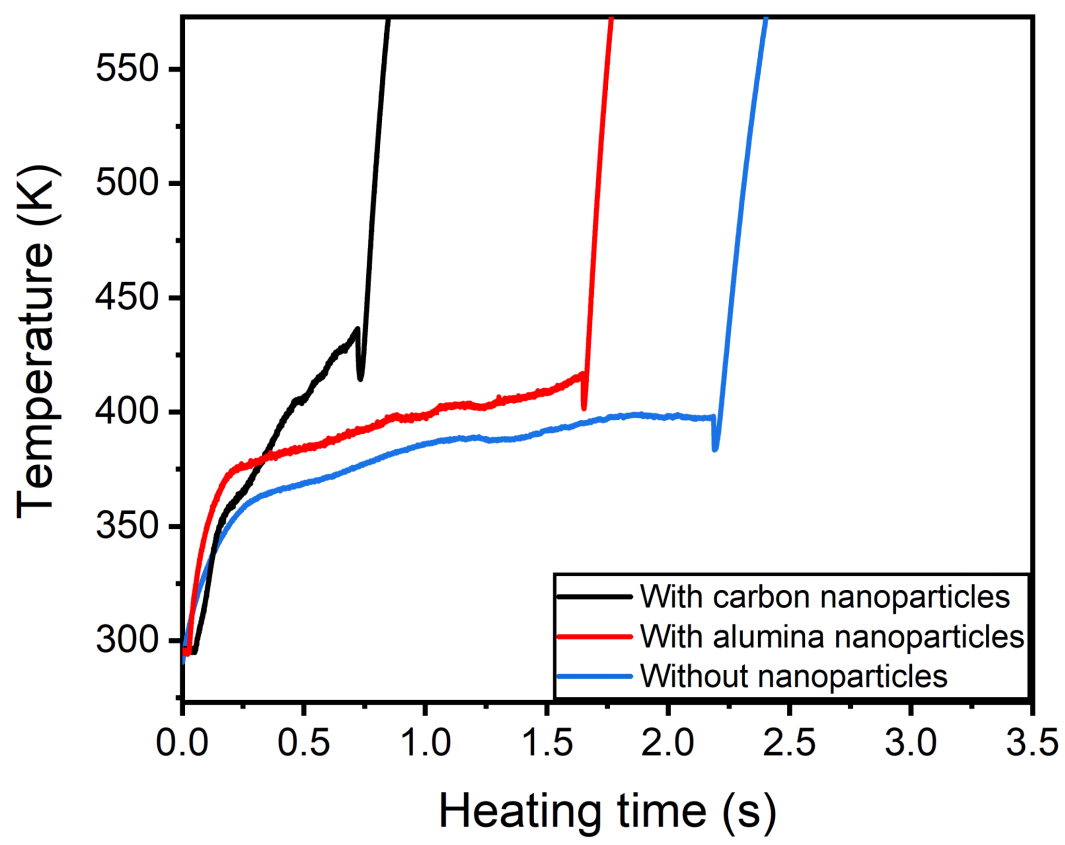

Figure 7. Temperature variation during droplet micro-explosion. The three curves correspond to the three micro-explosion processes shown in Figure 6.

By comparing the temperature change of the three droplets, it can be found that the temperature of the emulsion droplet with carbon nanoparticles rises rapidly to a high value. The temperature rise of the emulsion droplets with alumina nanoparticles is faster than that of the neat emulsion droplet, but much slower than that of the emulsion droplet with carbon nanoparticles. The emulsion droplet without nanoparticles has the lowest temperature and slowest temperature rise among the three emulsion droplets. This can be mainly attributed to the different absorption of infrared laser radiation energy by the emulsion droplets caused by nanoparticles. According to the absorbance in Figure 2, the infrared absorption of the emulsion droplet with carbon nanoparticles is much larger than that of the emulsion droplet with alumina nanoparticles. At the same time, carbon nanoparticles have higher thermal conductivity, as shown in Table 1, resulting in a faster rise in the temperature of the emulsion droplets. However, the absorbance of emulsion droplets with alumina nanoparticles to infrared radiation is slightly higher than that of neat emulsion droplets, so the amplitude and rate of temperature rise are also slightly higher than those of neat emulsion droplets.

The temperature drop at the instant of micro-explosion is an important feature of the micro-explosion process. Comparing the temperature drop of different emulsion droplets, we can see that the temperature drop of the emulsion droplets with carbon nanoparticles is the largest, while the temperature drop of the emulsion droplets without nanoparticles is the smallest. This is mainly because the emulsion droplets with carbon nanoparticles have the most severe micro-explosion and the most severe vaporization, thus absorbing a large amount of heat and bringing a significant reduction in the temperature of the emulsion droplets. For the emulsion droplets without nanoparticles, the micro-explosion is much weaker, so the absorbed heat is less, and the reduction of emulsion droplet temperature is also smaller.

From the high-speed images, we can see much mist can be produced during the micro-explosion process, as shown in Figure 6. Our previous studies have shown that the concentration of mist can reflect the micro-explosion intensity [41]. Therefore, to quantitatively characterize the mist concentration and micro-explosion intensity, a customized image processing program is used to identify the mist concentration. Here, the mist concentration is represented by the cumulated grey value (unitless) of the mist in high-speed images (see Section S2 in the Supplementary Material for the details of the image processing method).

The mist concentration curves of the three emulsion droplets during micro-explosion are shown in Figure 8. As the micro-explosion process proceeds, the mist concentration curve first rises and then declines. This is due to the vaporization of the dispersed phase inside the emulsion droplets at a high temperature. When there is enough vapor, the pressure inside the droplet is large. The internal vapor breaks through the oil film and causes micro-explosion of the droplets, resulting in a large amount of superheated vapor. Then the vapor is condensed by the low-temperature ambient gas to produce a large amount of mist. After that, due to the aerodynamic force, the mist spreads rapidly around, so the mist concentration curve declines rapidly. Comparing the concentration curves of the three kinds of emulsion droplets, we can see that there is much mist during the micro-explosion process of emulsion droplets with carbon nanoparticles. The peak value of the mist concentration curve is about several times that of the emulsion droplets with alumina nanoparticles. The peak value of the mist concentration curve of the emulsion droplets with alumina nanoparticles is similar to that of the pure emulsion droplets, which is less than the peak value of the emulsion droplets with carbon nanoparticles.

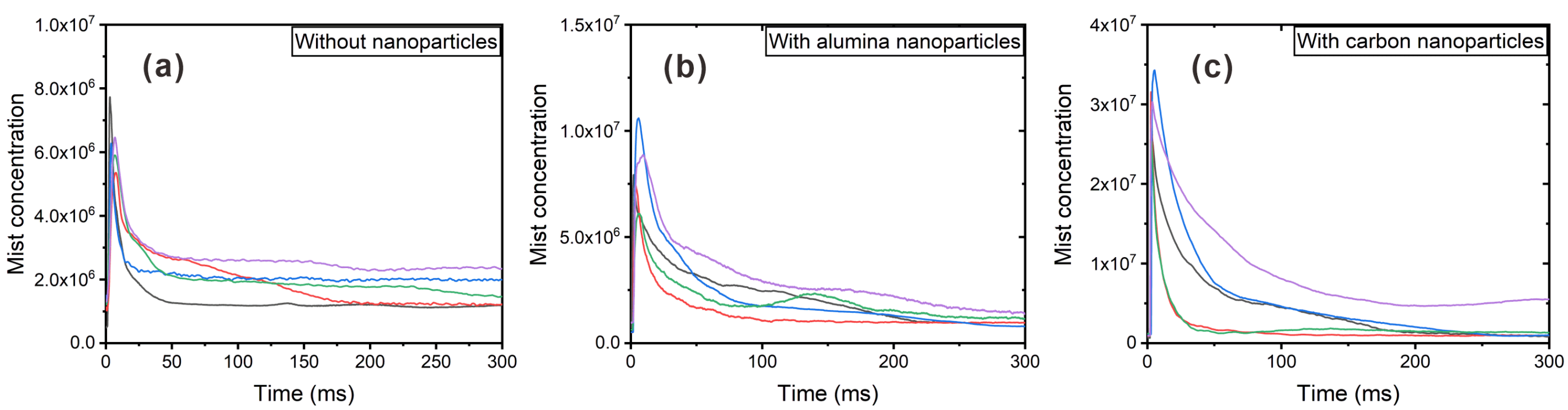

Figure 8. Mist concentration curve during intense micro-explosion of different emulsion droplets: (a) without nanoparticles, (b) with alumina nanoparticles, (c) with carbon nanoparticles. Here, $w_{NP}$ = 0.5 wt%, and $\phi_{water}$ = 30 vol%.

To further quantitatively describe the concentration change of mist during micro-explosion, we used two variables, $\Delta P$ and $T_{80}$. The concentration difference $\Delta P = P_{max} - P_{final}$ is used to quantify the peak of the mist concentration, where $P_{max}$ and $P_{final}$ are the maximum and final mist concentration, respectively. The duration $T_{80}$ is the period when $P > 0.8P_{max}$. The concentration difference $\Delta P$ and the duration $T_{80}$ of the micro-explosion processes are plotted on a map, as shown in Figure 9. It can be found that intense micro-explosion has large $\Delta P$ and small $T_{80}$, while weak micro-explosion has small $\Delta P$ and large $T_{80}$. For emulsion droplets with carbon nanoparticles, due to the violent vaporization in the heating process, intense micro-explosions occur, resulting in a large amount of mist. At the same time, the generated mist quickly disperses around, resulting in a very small $T_{80}$ and a very large $\Delta P$. Therefore, the micro-explosion results of carbon nanoparticles are mainly concentrated in the left-upper corner of the map. The micro-explosion results of alumina nanoparticles are not as severe as that of carbon nanoparticles, but it is more severe than that of emulsion droplets without nanoparticles. During the breakup process, dense mist can be generated and then disperses rapidly to the surroundings, resulting in a significant increase in $T_{80}$ compared with emulsion droplets with carbon nanoparticles, but it is smaller than that of emulsion droplets without nanoparticles. The curve $\Delta P=1.3\times10^{5}\cdot T_{80}$ can be used to distinguish between intense and weak micro-explosions, as shown by the dotted line in Figure 9.

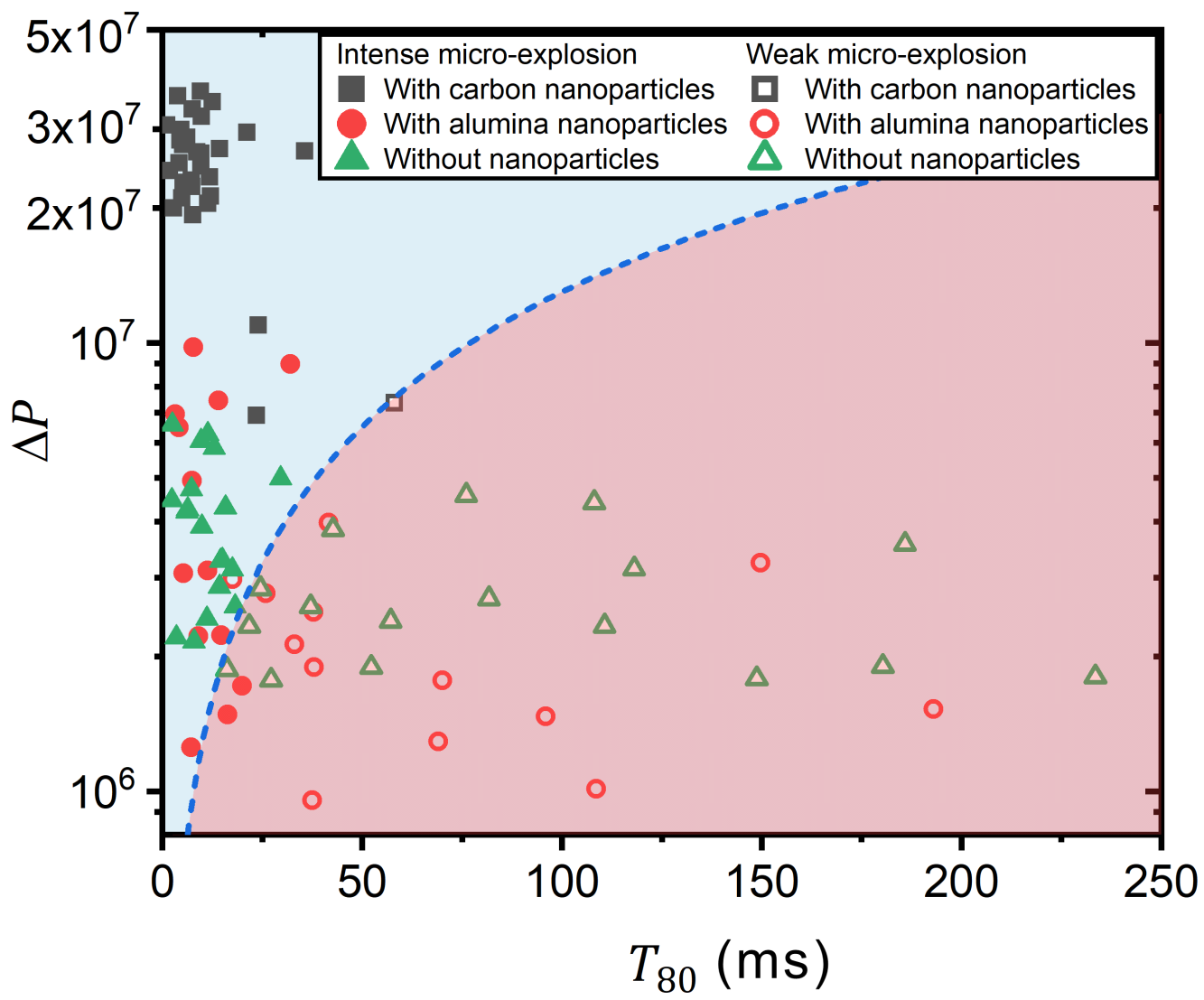

Figure 9. Map of different micro-explosion modes of different emulsion droplets. Here, $w_{\mathrm{NP}}$ = 0.5 wt%, and $\phi_{\mathrm{water}}$ = 30 vol%.

The probabilities of different outcomes (i.e., puffing, intense micro-explosion, and weak micro-explosion) are obtained by repeating the experiment of different emulsion droplets, as shown in Figure 10. The probability of intense micro-explosion is higher than that of neat emulsion droplets, while the probability of intense micro-explosion of emulsion droplets with carbon nanoparticles is even up to 97%, indicating that the addition of carbon nanoparticles almost always produces intense micro-explosion. This can be attributed to the large absorbance of the emulsion droplets with carbon nanoparticles. Under infrared laser radiation, the carbon nanoparticles increase the absorption of the radiation energy of the emulsion droplets, resulting in a significant increase in the droplet temperature during the heating process. Finally, the water phase inside the emulsion droplet is violently vaporized to produce a large amount of superheated vapor, forming an intense micro-explosion. In addition, from the evaporation phenomenon of emulsion droplets with nanoparticles observed under a microscope (as shown in Figure 4), it can be found that a large number of particle agglomerations occur in the later stage of evaporation, especially for carbon nanoparticles. The increase in droplet temperature further aggravates the agglomeration of particles. On one hand, high-temperature conditions aggravate the Brownian motion of particles, increasing the probability of particle collision. On the other hand, high temperature can increase the attraction (van der Waals Force) and reduce the repulsion (electric double layer force) between particles, resulting in particle agglomeration [42]. In the process of droplet heating, the dispersed water phase inside the emulsion droplets vaporizes violently to produce a large amount of superheated vapor. However, the agglomeration of particles can hinder the vapor diffusion in the droplet. Therefore, superheated vapor inside the emulsion droplets with carbon nanoparticles requires more energy to break the droplets, which increases the superheat of the water phase and the vapor pressure inside the droplets. Therefore, when the vapor breaks through the oil film, an intense micro-explosion will occur, resulting in the droplet being broken into many secondary droplets. For the emulsion droplets with alumina nanoparticles, the agglomeration of alumina particles is not as severe as that of carbon nanoparticles, so the breakup strength and the breakup temperature are not as high as that for carbon nanoparticles.

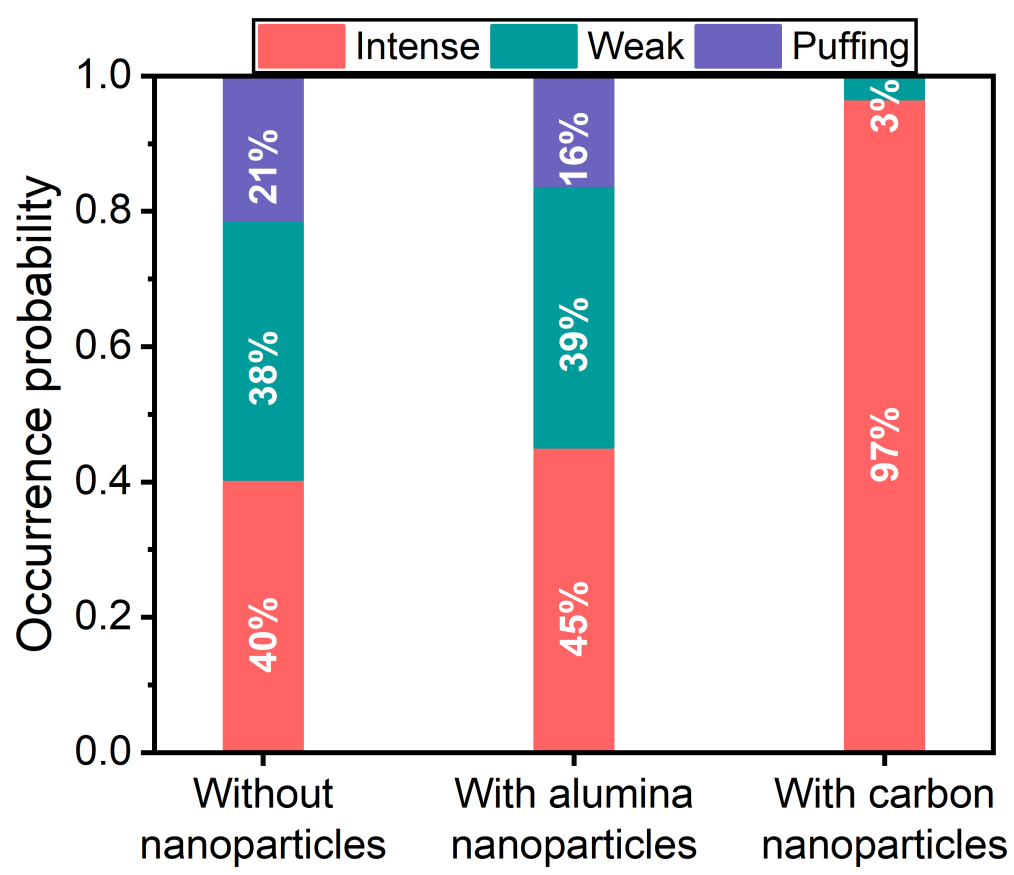

Figure 10. Effect of nanoparticles on the breakup mode. Here, $w_{NP}$ = 0.5 wt%, and $\phi_{water}$ = 30 vol%. The probability was calculated based on 30 groups of repeated experiments for each emulsion droplet composition.

### 3.2 Effect of nanoparticle concentration

To study the influence of particle concentration on the micro-explosion, alumina nanoparticles were selected to prepare emulsion droplets, and the mass fraction of alumina nanoparticles was changed from 0.05--3.0 wt%. The micro-explosion phenomenon at different mass fractions of alumina nanoparticles is shown in Figure 11. When the mass fraction of alumina is very low (0.05 wt%, as shown in Figure 11a), the fragmentation of the droplets during the breakup process is very weak, and puffing is observed. At the same condition, weak micro-explosion also occurs, producing large secondary droplets. When the mass fraction of alumina in the emulsion droplets increases to 0.5 wt%, due to the vaporization of water, the generated vapor is ejected from a certain position of the droplets, resulting in the local breakup of emulsion droplets, as shown in Figure 11b. A small amount of mist is observed, and the micro-explosion intensity is slightly enhanced. When the mass fraction of alumina in the emulsion droplets increases to 1.0--2.0 wt%, as shown in Figures 11c and d, the droplets expand instantaneously and then break up violently during the heating process. This is due to the violent vaporization of water, resulting in a large amount of vapor. The surface of the droplets is continuously stretched into a thin liquid film, and finally, the superheated vapor breaks through the oil film. Therefore, intense micro-explosion occurs and produces many tiny secondary droplets. At the same time, the superheated vapor is condensed into small droplets by the low ambient temperature, resulting in a large amount of mist. When the nanoparticle mass fraction continues to increase, as shown in Figure 11e, the breakup of the emulsion droplets is similar to that at the mass fraction of 2.0 wt%, and intense micro-explosion occurs, resulting in many secondary droplets, accompanied by much mist.

The occurrence probabilities of different breakup modes for emulsion droplets with different mass fractions of alumina nanoparticles are shown in Figure 12. When the nanoparticle mass fraction is less than 0.5 wt%, weak micro-explosion is the major mode. When the nanoparticles increase to about 1.0--3.0 wt%, the probability of intense micro-explosion is very high (i.e., above 80%), especially at the mass fraction of 2.0 wt%. This is because as the concentration of alumina nanoparticles increases, the alumina nanoparticles facilitate the absorption of radiation energy, inhibit the diffusion of superheated vapor, and ultimately promote micro-explosion. However, excessive nanoparticles also lead to more nucleation sites locally, resulting in local bubble breakup and release of vapor, which is not conducive to further enhancing micro-explosion. Therefore, the results show that at the mass fraction of about 2.0 wt%, alumina nanoparticles significantly enhance the breakup strength, and are conducive to fuel atomization.

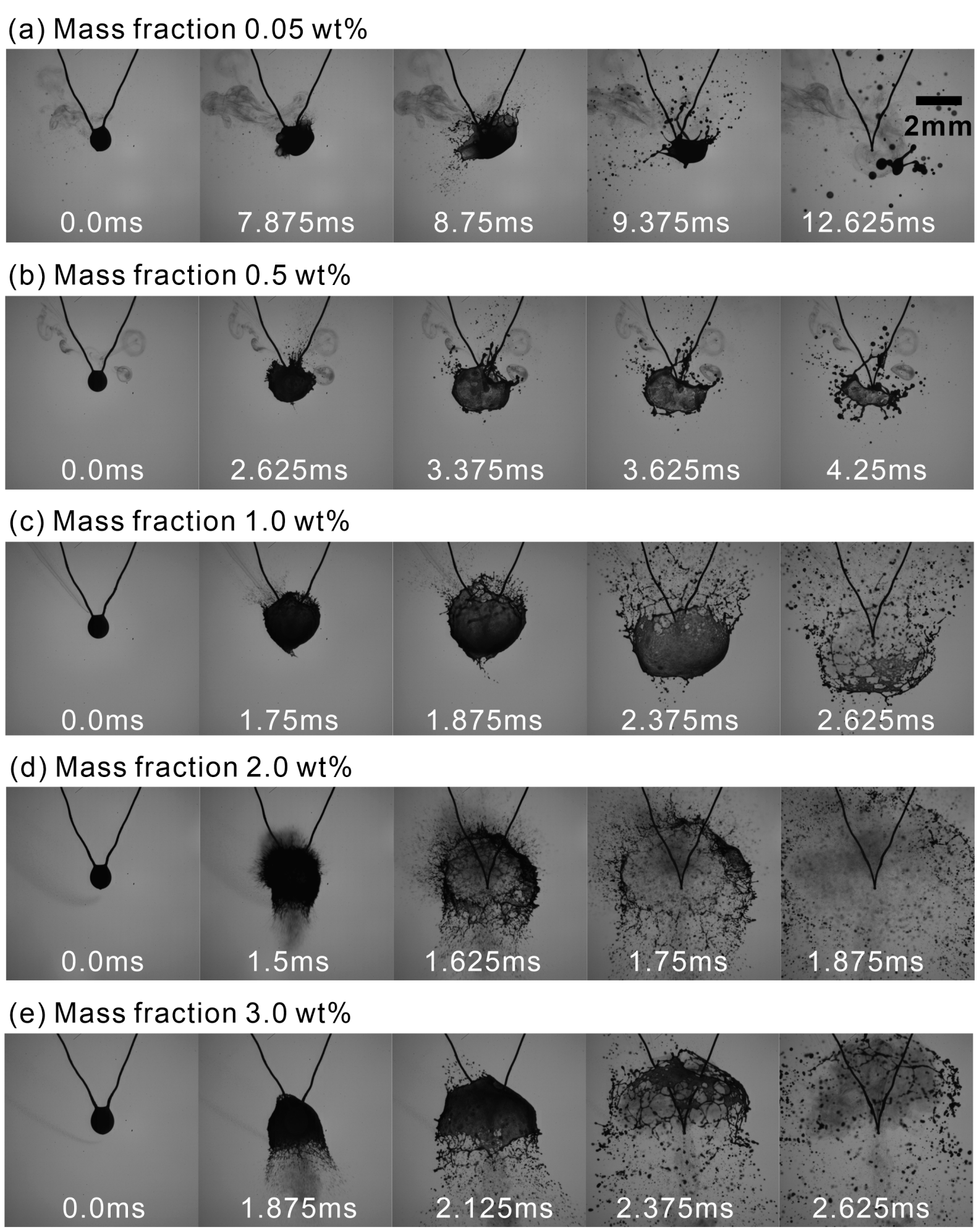

Figure 11. Micro-explosion of emulsion droplets with different alumina nanoparticle mass fractions ( $w_{NP}$ ): (a) 0.05 wt%, (b) 0.5 wt%, (c) 1.0 wt%, (d) 2.0 wt%, (e) 3.0 wt%. Here, $\phi_{water}$ = 30 vol%.

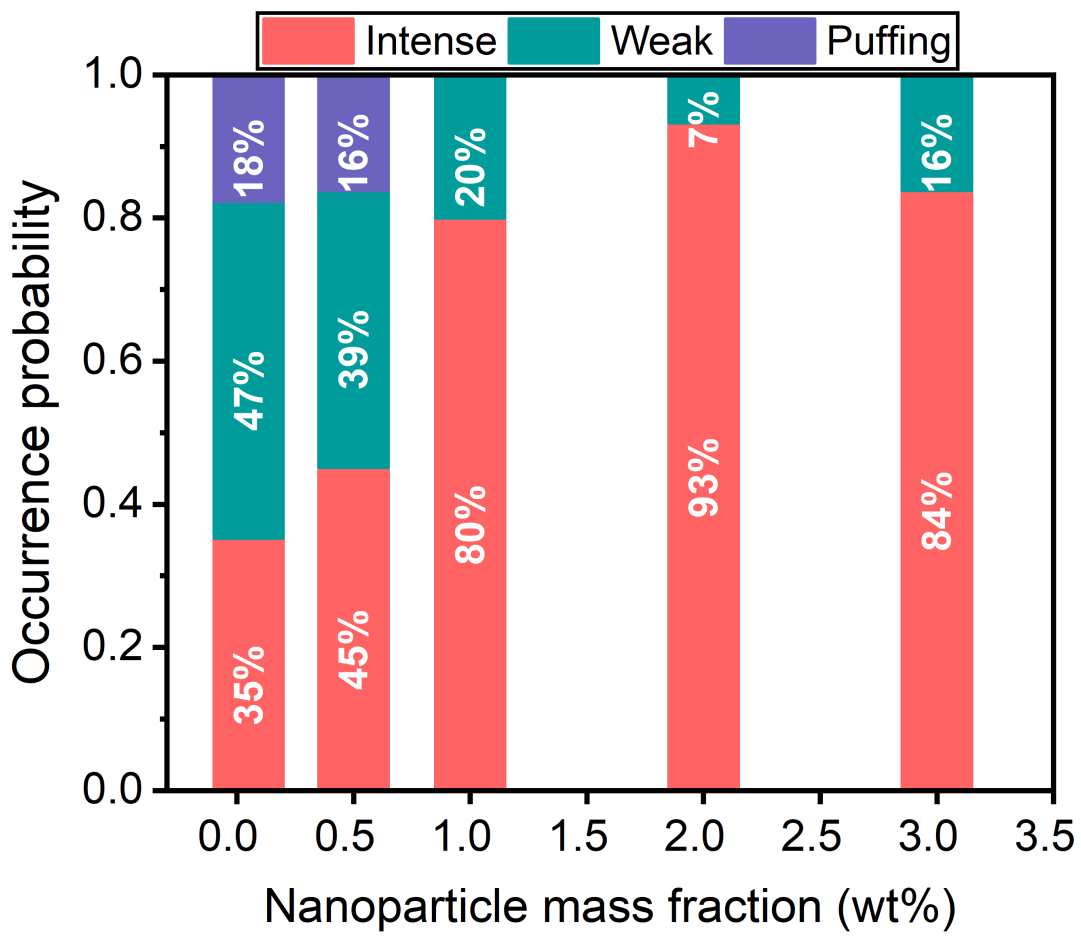

Figure 12. Effect of nanoparticle mass fraction on the breakup mode of emulsion droplets with alumina nanoparticles. Here, $\phi_{water}$ = 30 vol%. The probability was calculated based on 30 groups of repeated experiments for each nanoparticle mass fraction.

### 3.3 Effect of water content

Water, as a major content in the emulsion droplets, has a great influence on the micro-explosion process. The mass fraction of alumina nanoparticles is fixed at 1.0 wt%, and the water content was changed in the range of 10--40 vol%. The micro-explosion processes under different water content are shown in Figure 13. When the water content is low (10 vol%, as shown in Figure 13a), the puffing phenomenon occurs during the heating process, and the repeated puffing process produces a small number of secondary droplets. At the water content of 20 vol% (as shown in Figure 13b), many dispersed water droplets evaporate in the heating process. When the vapor pressure is large enough, the internal vapor can break the emulsion droplets, resulting in intense micro-explosion. The liquid film is gradually stretched, and then the ligaments break up into many secondary droplets. When the water content is increased to 30--40 vol%, the breakup strength does not further increase, and it produces intense micro-explosion accompanied by much mist.

The occurrence probability of different breakup modes under different water contents is shown in Figure 14. When the water content is low, only puffing and weak micro-explosion occur. When the water content increases to about 20 vol%, due to the water vaporization at high temperature, a large amount of vapor is generated and the vapor pressure is large enough to break through the droplet surface, resulting in intense micro-explosion. However, if further increasing the water content to 30--40 vol%, the probability of intense micro-explosion almost does not change. In summary, increasing the water content tends to promote micro-explosion, which is beneficial to the secondary atomization of the emulsion droplet with nanoparticles. However, when the water content inside the droplets exceeds 20 vol%, the influence of water content becomes weak. This is because, at high water content, the evaporation of water inside the emulsion droplets is maintained at about 20 vol%, consistent with our previous studies of emulsion droplets without nanoparticles [41,43].

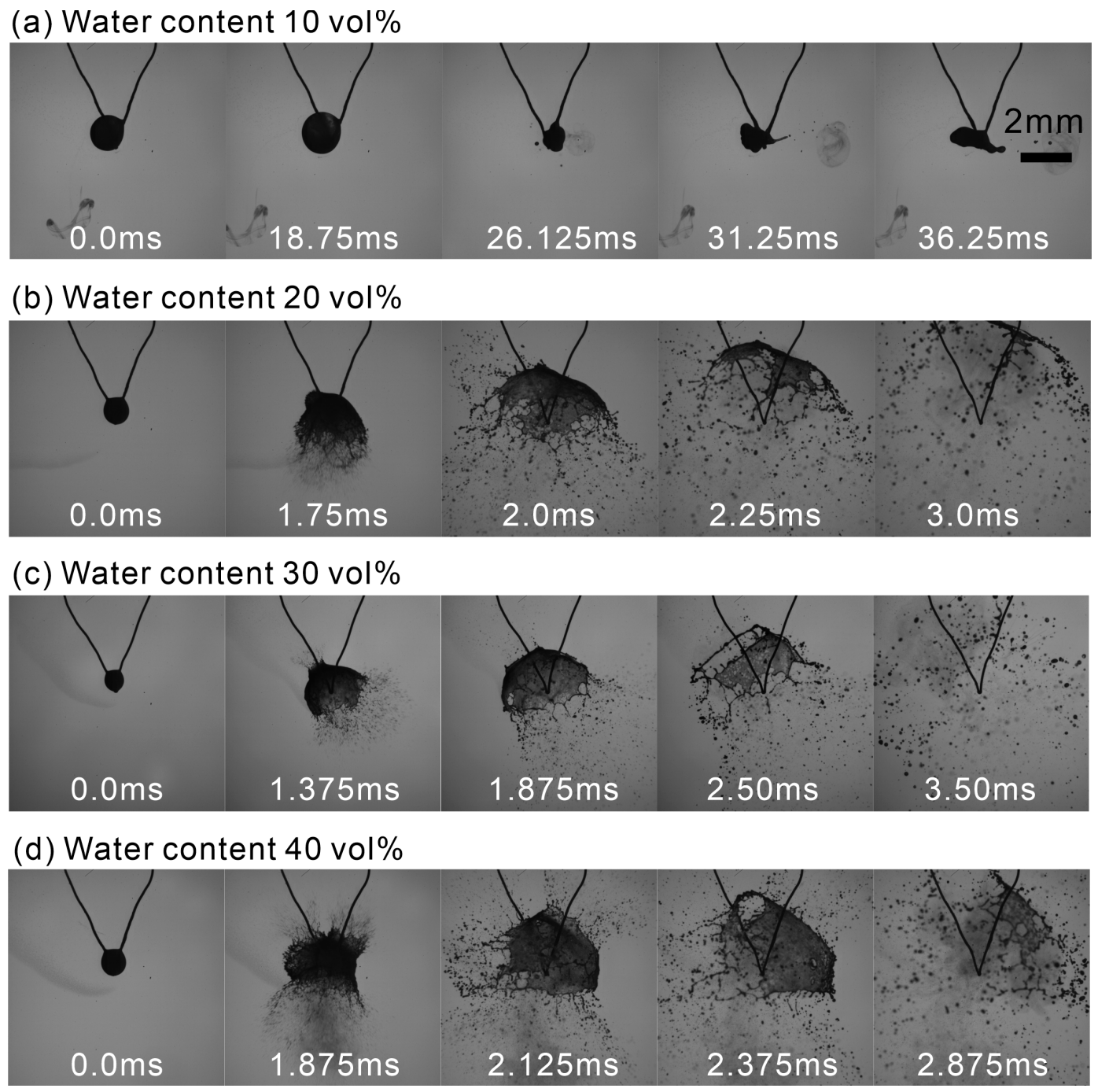

Figure 13. Micro-explosion of emulsion droplets with alumina nanoparticles, with different water contents ($\phi_{water}$): (a) 10 vol%, (b) 20 vol%, (c) 30 vol%, (d) 40 vol%. Here, $w_{NP}$ = 0.5 wt%.

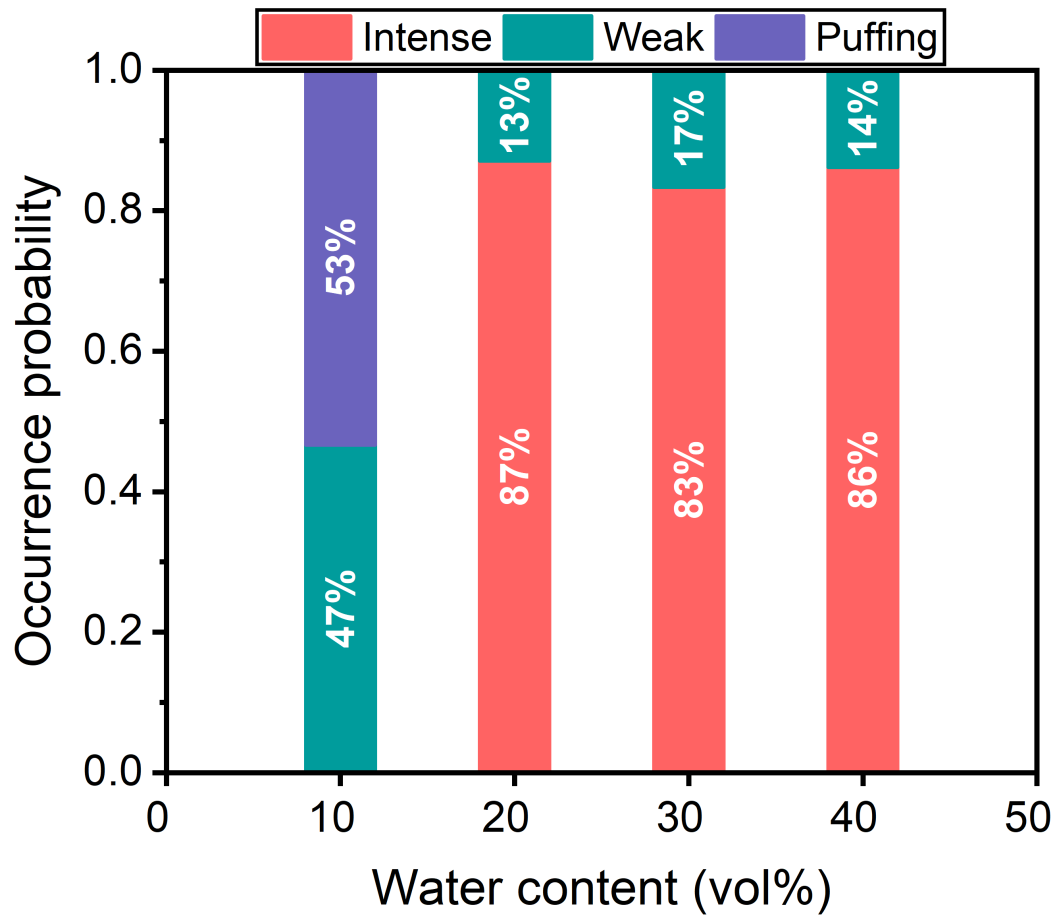

Figure 14. Effect of water content $\phi_{water}$ on the breakup mode of emulsion droplets with alumina nanoparticles. Here, $w_{NP}$ = 0.5 wt%. The probability was calculated based on 30 groups of repeated experiments for each water content.

## 4. Conclusions

In this paper, the micro-explosion of emulsion droplets with nanoparticles is investigated. The experimental results show that the presence of nanoparticles can greatly improve the strength and probability of micro-explosion, especially for carbon nanoparticles. This is mainly due to the large absorbance of the emulsion droplets with carbon nanoparticles and the increase in the absorption of radiation energy, which leads to intense vaporization of water inside emulsion droplets. In addition, many particle agglomerations occur in the later stage of evaporation, especially carbon nanoparticles form many agglomerations. The agglomeration of particles hinders the diffusion of vapor inside the emulsion droplet, so the superheated vapor inside the emulsion droplets requires more energy to break the emulsion droplets. Therefore, when the superheated vapor breaks through the oil film, intense micro-explosion can occur. The effects of nanoparticle mass fraction and water content are also studied. Increasing the nanoparticle mass fraction can increase the strength of micro-explosion. Alumina nanoparticles at the mass fraction of 2.0 wt% can significantly enhance the breakup strength, and are conducive to fuel atomization. This is mainly because of more energy absorption and more heterogenous nucleation sites. Increasing the water content also can increase the strength of micro-explosion when the water content is low. A water content of 20 wt% is remarkably beneficial for the secondary atomization of the emulsion droplet with nanoparticles.

This study mainly considers the effects of nanoparticles on the micro-explosion of emulsion droplets. Because of the complexity of the process, many aspects of the micro-explosion of emulsion droplets with nanoparticles are still unknown and need further studies. These include, but are not limited to, microscopic heat transfer of nanoparticles, nucleation mechanism of micro-explosion, macroscopic models and correlations of micro-explosion and its dependences on fuel compositions. Further tests in engine conditions are also helpful for the applications of emulsified fuels with nanoparticles.

## Acknowledgments

This work was supported by the National Natural Science Foundation of China (Grant Nos. 51976133, 51676137, and 51921004).

## References

[1] W.A. Sirignano, Fuel droplet vaporization and spray combustion theory, Progress in Energy and Combustion Science 9(4) (1983) 291-322.

[2] A. Williams, Combustion of liquid fuel sprays, Butterworth-Heinemann, 2013.

[3] Sangeeta, S. Moka, M. Pande, M. Rani, R. Gakhar, M. Sharma, J. Rani, A.N. Bhaskarwar, Alternative fuels: An overview of current trends and scope for future, Renewable and Sustainable Energy Reviews 32 (2014) 697-712.

[4] R.K. Gopidesi, P. Selvi Rajaram, A review on emulsified fuels and their application in diesel engine, International Journal of Ambient Energy 43(1) (2022) 732-740.

[5] P.K. Mondal, B.K. Mandal, A comprehensive review on the feasibility of using water emulsified diesel as a CI engine fuel, Fuel 237 (2019) 937-960.

[6] V.W. Khond, V.M. Kriplani, Effect of nanofluid additives on performances and emissions of emulsified diesel and biodiesel fueled stationary CI engine: A comprehensive review, Renewable and Sustainable Energy Reviews 59 (2016) 1338-1348.

[7] X. Zhang, T. Li, B. Wang, Y. Wei, Superheat limit and micro-explosion in droplets of hydrous ethanol-diesel emulsions at atmospheric pressure and diesel-like conditions, Energy 154 (2018) 535-543.

[8] G.S. Nyashina, K.Y. Vershinina, P.A. Strizhak, Impact of micro-explosive atomization of fuel droplets on relative performance indicators of their combustion, Fuel Processing Technology 201 (2020) 106334.
[9] S. Shen, K. Sun, Z. Che, T. Wang, M. Jia, J. Cai, Mechanism of micro-explosion of water-in-oil emulsified fuel droplet and its effect on soot generation, Energy 191 (2020) 116488.
[10] S.S. Sazhin, T. Bar-Kohany, Z. Nissar, D. Antonov, P.A. Strizhak, O.D. Rybdylova, A new approach to modelling micro-explosions in composite droplets, International Journal of Heat and Mass Transfer 161 (2020) 120238.
[11] S.S. Sazhin, O. Rybdylova, C. Crua, M. Heikal, M.A. Ismael, Z. Nissar, A.R.B.A. Aziz, A simple model for puffing/micro-explosions in water-fuel emulsion droplets, International Journal of Heat and Mass Transfer 131 (2019) 815-821.
[12] D.V. Antonov, R.M. Fedorenko, P.A. Strizhak, Characteristics of child droplets during micro-explosion and puffing of suspension fuel droplets, International Journal of Heat and Mass Transfer 209 (2023) 124106.
[13] P. Strizhak, R. Volkov, O. Moussa, D. Tarlet, J. Bellettre, Child droplets from micro-explosion of emulsion and immiscible two-component droplets, International Journal of Heat and Mass Transfer 169 (2021) 120931.
[14] A.B. Koc, M. Abdullah, Performance and NOx emissions of a diesel engine fueled with biodiesel-diesel-water nanoemulsions, Fuel Processing Technology 109 (2013) 70-77.
[15] H. Venu, V. Madhavan, Effect of $Al_2O_3$ nanoparticles in biodiesel-diesel-ethanol blends at various injection strategies: Performance, combustion and emission characteristics, Fuel 186 (2016) 176-189.
[16] S. Basu, A. Miglani, Combustion and heat transfer characteristics of nanofluid fuel droplets: A short review, International Journal of Heat and Mass Transfer 96 (2016) 482-503.
[17] H. Xie, H. Lee, W. Youn, M. Choi, Nanofluids containing multiwalled carbon nanotubes and their enhanced thermal conductivities, Journal of Applied Physics 94(8) (2003) 4967.
[18] C. Wu, T.J. Cho, J. Xu, D. Lee, B. Yang, M.R. Zachariah, Effect of nanoparticle clustering on the effective thermal conductivity of concentrated silica colloids, Physical Review E 81(1) (2010) 011406.
[19] N. Damanik, H.C. Ong, C.W. Tong, T.M.I. Mahlia, A.S. Silitonga, A review on the engine performance and exhaust emission characteristics of diesel engines fueled with biodiesel blends, Environmental Science and Pollution Research 25(16) (2018) 15307-15325.
[20] M.A. Lenin, M.R. Swaminathan, G. Kumaresan, Performance and emission characteristics of a DI diesel engine with a nanofuel additive, Fuel 109 (2013) 362-365.
[21] N. Sabet Sarvestany, A. Farzad, E. Ebrahimnia-Bajestan, M. Mir, Effects of magnetic nanofluid fuel combustion on the performance and emission characteristics, Journal of Dispersion Science and Technology 35(12) (2014) 1745-1750.
[22] A.C. Sajeevan, V. Sajith, Synthesis of stable cerium zirconium oxide nanoparticle – Diesel suspension and investigation of its effects on diesel properties and smoke, Fuel 183 (2016) 155-163.

[23] D. Mei, X. Li, Q. Wu, P. Sun, Role of cerium oxide nanoparticles as diesel additives in combustion efficiency improvements and emission reduction, Journal of Energy Engineering 142(4) (2016) 04015050.

[24] J. Sadhik Basha, R.B. Anand, The influence of nano additive blended biodiesel fuels on the working characteristics of a diesel engine, Journal of the Brazilian Society of Mechanical Sciences and Engineering 35(3) (2013) 257-264.

[25] S. Vellaiyan, C.M.A. Partheeban, Combined effect of water emulsion and ZnO nanoparticle on emissions pattern of soybean biodiesel fuelled diesel engine, Renewable Energy 149 (2020) 1157-1166.

[26] S. Vellaiyan, Enhancement in combustion, performance, and emission characteristics of a biodiesel-fueled diesel engine by using water emulsion and nanoadditive, Renewable Energy 145 (2020) 2108-2120.

[27] S. Tanvir, L. Qiao, Droplet burning rate enhancement of ethanol with the addition of graphite nanoparticles: Influence of radiation absorption, Combustion and Flame 166 (2016) 34-44.

[28] J. Wang, X. Qiao, D. Ju, L. Wang, C. Sun, Experimental study on the evaporation and micro-explosion characteristics of nanofuel droplet at dilute concentrations, Energy 183 (2019) 149-159.

[29] B. Pathak, S. Basu, Phenomenology of break-up modes in contact free externally heated nanoparticle laden fuel droplets, Physics of Fluids 28(12) (2016) 123302.

[30] Z. Wang, B. Yuan, Y. Huang, J. Cao, Y. Wang, X. Cheng, Progress in experimental investigations on evaporation characteristics of a fuel droplet, Fuel Processing Technology 231 (2022) 107243.

[31] E. Khalife, M. Tabatabaei, A. Demirbas, M. Aghbashlo, Impacts of additives on performance and emission characteristics of diesel engines during steady state operation, Progress in Energy and Combustion Science 59 (2017) 32-78.

[32] X. Wang, J. Zhang, Y. Ma, G. Wang, J. Han, M. Dai, Z.Y. Sun, A comprehensive review on the properties of nanofluid fuel and its additive effects to compression ignition engines, Applied Surface Science 504 (2020) 144581.

[33] X. Wang, M. Dai, J. Wang, Y. Xie, G. Ren, G. Jiang, Effect of ceria concentration on the evaporation characteristics of diesel fuel droplets, Fuel 236 (2019) 1577-1585.

[34] B. Ashok, K. Nanthagopal, A. Mohan, A. Johny, A. Tamilarasu, Comparative analysis on the effect of zinc oxide and ethanox as additives with biodiesel in CI engine, Energy 140 (2017) 352-364.

[35] S. Tanvir, S. Biswas, L. Qiao, Evaporation characteristics of ethanol droplets containing graphite nanoparticles under infrared radiation, International Journal of Heat and Mass Transfer 114 (2017) 541-549.

[36] I.A. S. Ferrão, A.R.R. Silva, A.S.O.H. Moita, M.A.A. Mendes, M.M.G. Costa, Combustion characteristics of a single droplet of hydroprocessed vegetable oil blended with aluminum nanoparticles in a drop tube furnace, Fuel 302 (2021) 121160.

[37] M. Ghamari, A. Ratner, Combustion characteristics of colloidal droplets of jet fuel and carbon based nanoparticles, Fuel 188 (2017) 182-189.

[38] B. Dhinesh, M. Annamalai, A study on performance, combustion and emission behaviour of diesel engine powered by novel nano nerium oleander biofuel, Journal of Cleaner Production 196 (2018) 74-83.

[39] J.S. Basha, R.B. Anand, An experimental investigation in a diesel engine using carbon nanotubes blended water–diesel emulsion fuel, Proceedings of the Institution of Mechanical Engineers, Part A: Journal of Power and Energy 225(3) (2011) 279-288.
[40] J. E, Z. Zhang, J. Chen, M. Pham, X. Zhao, Q. Peng, B. Zhang, Z. Yin, Performance and emission evaluation of a marine diesel engine fueled by water biodiesel-diesel emulsion blends with a fuel additive of a cerium oxide nanoparticle, Energy Conversion and Management 169 (2018) 194-205.
[41] H. Zhang, Z. Lu, T. Wang, Z. Che, Mist formation during micro-explosion of emulsion droplets, Fuel 339 (2023) 127350.
[42] Z. Yan, X. Huang, C. Yang, Deposition of colloidal particles in a microchannel at elevated temperatures, Microfluidics and nanofluidics 18(3) (2014) 403-414.
[43] S. Shen, K. Sun, Z. Che, T. Wang, M. Jia, J. Cai, An experimental investigation of the heating behaviors of droplets of emulsified fuels at high temperature, Applied Thermal Engineering 161 (2019) 114059.